\documentclass[12pt]{iopart}

\usepackage{graphicx}

\begin{document}

\title[]{Comparison between classical potentials and ab initio for silicon 
under large shear}

\author{J. Godet, L. Pizzagalli, S. Brochard, P. Beauchamp}

\address{Laboratoire de M\'etallurgie Physique, CNRS UMR 6630, Universit\'e 
de Poitiers,  B.P. 30179, 86962 Futuroscope Chasseneuil Cedex, France}

\ead{Laurent.Pizzagalli@univ-poitiers.fr}

\begin{abstract}

The homogeneous shear of the \{111\} planes along the $<110>$ direction of 
bulk silicon  has been investigated using ab initio techniques, to better 
understand the strain properties of both shuffle and glide set planes. Similar 
calculations have been done with three empirical potentials, 
Stillinger-Weber, Tersoff and EDIP, in order to find the one 
giving the best results under large shear strains. 
The generalized stacking fault energies have also been calculated with 
these potentials for complementing this study. It turns out that the 
Stillinger-Weber potential better reproduces the ab initio results, 
for the smoothness and the amplitude of the energy variation as well 
as the localisation of shear in the shuffle set.

\end{abstract} 

\submitto{\JPCM} 
\maketitle

\section{Introduction}

Dislocations in materials generate a long-range strain field, which must be 
taken into account in atomistic simulations for a proper treatment of the 
defects. As a consequence, large systems including thousands of atoms have 
to be employed. Only in specific cases it is possible to use a limited 
number of atoms. For example, if the considered dislocations are ideally 
straight and infinite, the core structure may be investigated with few 
hundreds of atoms. Precise electronic and atomic structure calculations can 
then be performed using ab initio methods 
\cite{Obe95PRB,Jus00JPCM,Ewe01JPCM,Jus01SSC,Piz03PM}. Still, investigating 
the formation, mobility or interaction of dislocations requires a 
large system with a few thousands of atoms \cite{Ras00PMA}, preventing the 
use of such methods. Then the atomic simulation must be performed with 
empirical interatomic potentials.

With such potentials, the simulations can be performed with a large number 
of atoms and for a long time scale, with a moderate cost in calculation 
time. Then, they are a valuable mean for simulating the activation of 
complex physical mechanisms, using techniques like molecular dynamics. 
However, it remains difficult to evaluate the validity of the results 
obtained from such studies. Interatomic potentials are built in order to 
reproduce with a fair accuracy a limited number of physical quantities, 
usually at equilibrium. The reliability of potentials is then doubtful when 
one investigates physical mechanisms or configurations where the atomic 
structure is far from its equilibrium state, such as, for example, 
reconstruction of surfaces, point or extended defects. Moreover, the nature 
of material is also an important factor. It is commonly agreed that the 
results obtained using interatomic potentials are qualitatively better for 
metals than in the case of covalent materials. This problem is very 
sensible for silicon in particular, since it is largely studied, for 
technology purposes, or as a model for semiconductors. Several kinds of 
potentials have been proposed but it is difficult to assess the superiority 
of one over the other. In particular, the transferability is poor in many 
cases. Several comparative studies on elastic constants, bulk point 
defects, core properties of partial dislocation, structure of disordered 
phases have already been performed with these potentials 
\cite{Bal92PRB,Jus98PRB,Mor98JJAP}. They conclude that, for each kind of 
system or physical mechanism, one must determine the best interatomic 
potential.

Recently, Godet {\it et al} \cite{God02SM2} have studied the nucleation of dislocations from a 
surface step on silicon. The 
dislocation formation with this mechanism would explain the presence of 
dislocation observed in nano-materials \cite{Wu01PMA} where the dimensions 
are too small to allow a classical multiplication mechanism like Franck-
Read sources. This would also explain the appearance of dislocations from 
cleavage ledges, when silicon is plastically deformed at low temperature 
\cite{Arg01SM}. Since the observation of the very first stage of 
dislocation nucleation is difficult, the atomic simulation may bring up an 
interesting alternative. However, recently, Godet {\it et al} have shown 
that the obtained results were potential dependent \cite{God03UNP}. During the process of 
nucleation of the dislocations, the atomic structure is so largely deformed 
that the potentials go out of their usual domain of validity, which may 
explain these disagreements. It would be helpful to find out which is the 
silicon potential giving the best results in the case of large shears.

In this paper, an attempt is presented in view of determining the best 
interatomic potential for silicon submitted to large shear strain, by 
comparing results obtained for different potentials with ab initio 
calculations. First, considering homogeneous shear of bulk silicon, two 
criteria have been used for the potential selection. The first one bears 
upon the variation of the bulk energy  as a function of the applied strain. 
The second criterion is related to atomic configurations in that it 
considers how imposed atomic displacements distribute between the shuffle 
and the glide sets of the \{111\} planes. In particular, we focus on the 
mechanism of atomic bonds switching from one neighbor to another. After the 
homogeneous shears, a second part is devoted to generalized stacking fault 
(GSF) energy surfaces, in particular their shape in both shuffle and glide 
set planes at large fault vectors. All these situations contribute to a 
better understanding of the strain properties of both glide and shuffle 
planes, related to the mechanisms of dislocations nucleation and mobility 
\cite{Due96PML,Jus01PHY,Bul01PMA,Rab01MSE} or to the mechanisms of 
cleavage and fracture of this crystal \cite{Per00AM,Gal01PMA,Pir01PMA}.  

\section{Methodology} \subsection{Shear technics}

In ambient conditions silicon crystallizes into the  diamond cubic 
structure which is formed of two interpenetrated face centered cubic 
sublattices called in the following sublattice 1 and 2. In this 
structure, the dislocations glide in the \{111\} dense planes with a $1/2 
<110>$ Burgers vector corresponding to the shortest vector of the fcc 
lattice. We have studied the silicon bulk under large shear strain, along 
the \{111\} planes in the $<110>$ direction, from zero up to a given shear 
strain allowing to recover the cubic diamond structure. As the cubic 
diamond structure includes two fcc sublattices, there are two kinds of 
\{111\} planes, that are alternately piled up, the narrowly spaced between 
the (111) planes of the sublattices 1 and 2, and the widely spaced between 
2 and 1, called glide and shuffle set planes respectively \cite{Hir82WIL} 
(figure~\ref{strain}). To deform the crystal, we have progressively applied 
shear strain increments about 4\% on both sublattices 1 and 2, up to a 
shear strain called $\gamma_{23tot}\simeq 122$\%. This limit corresponds to 
the ratio of a slip in a shuffle set equal to a Burgers vector {\bf b} 
(={\bf x$_3$}) by the shuffle and glide set interplanar distance ($={\bf 
x_2}/3$), i.e. $\frac{x_3}{x_2/3} = \sqrt{3}/\sqrt{2}$. 
After each shear increment, the atomic positions belonging to the 
sublattice 2 are relaxed in all directions to minimize the energy. Note 
that the calculations are performed at 0 K and constant volume. The aim is 
to monitor the energy evolution during the shear, and determine how the 
homogeneous shear strain is divided between the shuffle set and the glide 
set. In this paper, we call displacement in one set (shuffle or glide), 
the shift after application of the strain (before relaxation), and shear 
strain in one set the ratio of the displacement by the interplanar 
distance in this set (figure~\ref{strain}). Note that, interplanar distances 
may vary because the sublattice 2 is free. However, as the shear is done at 
constant volume, the addition of the shuffle and glide set interplanar 
distances remains constant. For the same reason, the bulk shear stress 
$\sigma_{23}$ is obtained by the derivation of the atomic energy curve 
against the applied shear strain $\gamma_{23}$ .

\

We have also calculated the GSF energy and the restoring forces along the 
slip directions in shuffle and glide planes. The unrelaxed GSF energy 
surface is obtained by simply moving one half of the crystal rigidly with 
respect to the other half along a cut plane in the middle of the crystal. 
The GSF energy is defined as a function of the relative displacement {\bf 
f} of the two atomic planes immediately adjacent to the crystal cut plane. 
To calculate the GSF energy with atomic and volume relaxation, the atoms in 
the two planes immediately adjacent to the cut plane are restricted to move 
along the $<111>$ direction only, in order to keep the relative 
displacement {\bf f}, whereas the other atoms relax in all directions. 
Therefore, the actual relative displacement {\bf f} might be different from 
the displacement between the centers of both halves of the crystal. The 
restoring force in a given direction, corresponds to the derivative of the 
GSF energy versus the displacement {\bf f}. Here, we focused on the 
$<110>$ direction in the shuffle and glide set, and also on the $<112>$ 
direction in the glide set, since perfect dislocations can be dissociated 
in two Shockley partial dislocations with $1/6 <112>$ Burgers vectors in 
that set. The local shear stress required to maintain the displacement {\bf 
f} in both sides of the cut plane is directly proportional to the opposite 
of the restoring forces.

\subsection{Computational Methods}

First principles calculations of the bulk shear are 
performed using the ABINIT package \cite{Abinit}, the 
exchange correlation energy being determined within the 
local density approximation with the Teter Pade 
parametrization \cite{Goe96PRB} which reproduces Perdew-
Wang. The valence electron  wave functions are expanded in a 
plane wave basis with a cut off energy of 15 Ha. The ionic 
potential is modeled by a norm conserving pseudo-potential 
from Troullier and Martins \cite{Tro91PRB}. To simulate the 
bulk shearing  process, a periodic cell orientated along 
$1/2[121]$ ({\bf x$_1$}), $[\bar11\bar1]$ ({\bf x$_2$}) and 
$1/2[\bar101]$ ({\bf x$_3$}) is used, including 12 atoms, 
i.e. 6 \{111\} atomic planes (3 glide set and 3 shuffle set 
planes) along {\bf x$_2$}. For the reciprocal space 
integration, we have used 9 special k-points in the 
irreducible Brillouin zone when the cell is not sheared, and 
15 special k-points when the cell is sheared owing to the 
reduced symmetry. The k-points lattice obtained with the 
Monkhorst and Pack scheme \cite{Mon76PRB}, is the reciprocal 
of the super-lattice defined by the supercell in the real 
space by 3{\bf x$_1$}, 2{\bf x$_2$} and 5{\bf x$_3$}, the 
origin of this k-point lattice being shifted by a 
$[0.5,0.5,0.5]$ vector. The SCF cycle is stopped when the 
difference on total energy between two successives cycles is 
smaller than $10^{-10}$ Ha. Metallic occupation of levels is 
allowed using the Fermi-Dirac smearing occupation scheme. 
The atomic positions are relaxed using the Broyden  -
Fletcher - Goldfarb - Shanno minimization down to forces 
smaller than $5\times10^{-5}$ Ha/Bohr ($2.5\times10^{-3}$ 
eV/\AA).
We have compared our 
results with the ab initio study performed by Umeno {\it et al} 
\cite{Ume02MSE} where the same calculation is realized with a full 
relaxation of volume and ionic position but only up to 35\% of strain. 

For the empirical bulk shear calculations, three different interatomic 
potentials have been used, Stillinger-Weber (SW) \cite{Sti85PRB}, Tersoff 
\cite{Ter89PRB} and the Environment-Dependent Interatomic Potential (EDIP) 
\cite{Baz97PRB}. The pioneering potential of SW has only eight parameters 
and is fitted to few experimental properties of  both crystallized (cubic 
diamond) and liquid silicon. It consists of a linear combination of two and 
three body terms. The Tersoff functional form is fundamentally different 
from the SW form in that it includes many body interactions thanks to a 
bond order term. As a result, the strength of individual bonds is affected 
by the presence of surrounding atoms. The final version called T3 has 
eleven adjustable parameters fitted to ab initio results for several Si 
polytypes. The third potential, EDIP, has a functional form similar to that 
of Tersoff but slightly more complicated. It incorporates several 
coordination-dependent functions to adapt the interactions for different 
coordinations. 13 parameters are determined by fitting to a fairly small ab 
initio database. 

The dimensions of the calculation cell must be twice larger than the cut 
off radius of the interatomic potentials to minimize interaction of atoms 
with their images in neighboring cells. 
So a calculation cell with the 
same geometry as before is used, but containing 576 atoms. 
The relaxation of atomic positions is performed with a conjugate gradients 
algorithm until the magnitudes of the forces are smaller than $10^{-4}$ 
eV/\AA.

The GSF energy surfaces calculations have been performed with the three 
interatomic potentials and we have compared our results with the GSF energy 
surfaces obtained with first principles calculations by Juan and Kaxiras 
\cite{Kax93PRL2,Jua96PMA}. Note that several energy curves or unstable 
stacking fault energies have been calculated with those interatomic 
potentials \cite{Jus98PRB,Due91EMRS,Kon98PRB}. Here, periodic conditions 
are applied only along the $<112>$ and $<110>$ directions. In the third 
direction, the number of \{111\} planes is large enough (30) to avoid 
spurious interactions between the free surfaces and the crystal cut 
plane. The system  contains 1440 atoms.

\section{Results / discussion} 

\subsection{Homogeneous shear strains}

The calculation of homogeneous shear strain of bulk silicon has been 
performed ab initio and with the three interatomic potentials. The 
observation of the sheared atomic structure obtained with the ab initio 
calculation (figure~\ref{snapshot}) shows that the strains are essentially 
located in the shuffle set. The bonds between atoms across the shuffle set 
plane are successively weaken, broken and then formed again. This  is 
confirmed by the monitoring of the electronic density where the covalent 
character of the bonds progressively vanishes to reach a metallic character 
at half of the applied shear strain in the shear direction ($<110>$). Our 
calculation is in agreement with the ab initio study realized by Umeno {\it 
et al} \cite{Ume02MSE}, where it is found that the band gap is 
progressively closed with the applied shear. At the maximum of the applied 
shear strain, each shuffle set plane have been shifted by a Burgers vector 
of a perfect dislocation and the diamond crystal is recovered.

To compare the different interatomic potentials, the atomic energy  and the 
corresponding shear stresses as a function of the applied shear strain are 
calculated and represented in figure~\ref{shear}. For small strains, all 
the energy curves coincide and the stress curves are linear, indicating that 
the empirical potentials are fairly well fitted to the elastic 
coefficients. The shear modulus associated to $<110>$\{111\} shear at 
constant volume obtained from the ab initio calculation is around 52 GPa 
close to the value calculated with volume relaxation \cite{Ume02MSE} and 
also relatively close to the value obtained from the elastic coefficient 
calculated at 0 K ($C_{12}- \frac{1}{3}(2C_{44}+C_{12}- C_{11})$ = 48.3 
GPa) \cite{Kar97JPCM}. For larger strains, the potentials may be classified 
in two groups depending on whether they are close or not to ab initio. SW 
belongs to the first group with energy curves in fair agreement with ab 
initio which presents smooth maxima of similar heights at half of the applied shear, 
whereas EDIP and Tersoff are in the other group with larger maxima and 
angle-shaped curves. Regarding stresses, the variations of SW and ab initio 
are relatively smooth compared to EDIP and Tersoff, with similar 
theoretical shear strengths reached at about one fourth of the total applied 
strain, while the ones obtained with EDIP and Tersoff are larger and reached at  
about half of the applied strain (table~\ref{theo_stress}). Note that our ab 
initio theoretical shear strength at constant volume is relatively close to 
the value calculated with volume relaxation \cite{Ume02MSE}. SW seems to be 
the best interatomic potential to model the  shear stress evolution during 
the atomic bond switching. 
Probably the introduction of temperature would smooth the 
energy curves, and in the case of the Tersoff potential, 
would allow the crossing of the energy barrier to recover 
the diamond crystal. However, the general shape of 
calculated curves will be preserved, in particular for 
deformations corresponding to theoretical shear strenghs.

To analyze the atomic structure, the displacements and shear strains in 
both shuffle and glide set planes along the $<110>$ shear directions have 
been determined (figure~\ref{disp}). For applied strains up to half maximum, 
all potentials show a similar behavior, the displacements in the shuffle 
set plane following the applied strains, while those in the glide set 
oscillate weakly with a magnitude lower than 0.15 \AA. For larger applied strains, 
the displacements in the shuffle set reach the Burgers vector of a perfect 
dislocation. In the glide set they return to zero, except for the Tersoff 
potential where the displacements in the shuffle set remain practically 
constant and where the displacements are then located in the glide set 
along $<112>$. The variations of shear strains in both planes, show that 
the ab initio, SW and EDIP results are relatively close to each other. The 
interatomic potentials modeling properly these effects are SW and EDIP.

In figure~\ref{shear}, the shear strains of the glide set planes are 
represented with dashed line next to the bulk shear stress with full line. 
In all cases, while most of the displacements are localized in the shuffle 
set, the shear strains of the glide set are approximately linear with the 
bulk shear stresses, with a large shear modulus ($\mu$), for example 134~GPa 
with ab initio calculations. The strains localized in the glide set remain then 
elastic and linear whereas those of the shuffle set do not. Moreover the 
large shear modulus of the glide set shows that the displacements in the 
glide set, although always small, play an important role on the bulk shear 
stress. This is confirmed by the study of Umeno \cite{Ume02MSE} where it is 
concluded that the subtle displacements in the glide set have a remarkable 
effect on the shear stress.

\subsection{GSF energy and restoring force}

We have investigated the GSF energy surfaces and the  corresponding 
restoring forces in directions of Burgers vectors \textbf{b}, calculated with ab 
initio techniques \cite{Jua96PMA} and interatomic potentials, in order to compare the 
localized shear stresses in the shuffle and glide set planes. Two  
directions have been investigated $<110>$ in the shuffle and glide set for 
perfect dislocations, and $<112>$ in the glide set for Shockley partial 
dislocations. 

Usually, one considers the maxima of the GSF energy, i.e. the unstable 
stacking fault energy $\gamma_{us}$ (table~\ref{GSF}), as an important 
parameter for gliding. In addition with $\gamma_{us}$, we also determine 
the maxima of the restoring force, $\tau_{max}$, along the three 
directions (table~\ref{taumax}). In all cases, the lowest values are 
obtained for the $<110>$ direction in the shuffle set plane, as expected. The best 
$\gamma_{us}$ are given by Tersoff and EDIP, the SW potential tending to 
underestimate in the shuffle set and to overestimate in the glide set. 
Regarding the restoring force, the SW potential yields the best results, 
the large values for EDIP and Tersoff coming from the singularities in the 
curves. It appears that the sole determination of these maxima is not 
enough to discriminate between the potentials. An additional indication is 
given by the location of the maxima. The best agreement with ab initio is 
then obtained for the SW potential, with maxima in the vicinity of 0.3b, 
whereas for both EDIP and Tersoff, they are located at displacements greater 
than 0.4b. 

Instead of considering only maxima, we compared directly the variations. 
Along the three directions, the ab initio GSF energy variations, 
calculated by Juan et al \cite{Jua96PMA}, are smooth, with sinusoidal-shaped 
curves. Comparing with the GSF energy for the three potentials 
(figure~\ref{tau}), the best qualitative agreement is obtained with the SW 
potential. Both EDIP and Tersoff show large and abrupt variations of the 
GSF energy, as soon as the displacement is greater than 0.4b. In 
particular, distorted shapes and angular points are present for EDIP in 
the glide set, and Tersoff in the shuffle set. If we focus on the favored 
glide direction for perfect dislocation, i.e. the $<110>$ in the shuffle 
set plane, it appears that the Tersoff potential shows the worst results, 
with a deep local minimum at 0.5b. Using EDIP and SW, instead, leads to an 
energy maximum at 0.5b, as obtained with the ab initio calculation. The whole EDIP and SW GSF energy 
curves are in fair agreement with ab initio, although the smoothest 
variations are obtained with SW.

More indications can be gained from the calculation of the restoring 
forces in the three cases. The variations, represented in 
figure~\ref{tau}, are similar to the bulk shear 
stress curves, shown in figure~\ref{shear}. The various conclusions drawn from the analysis of 
the GSF energy variations remain valid here. The best agreement is 
obtained for the SW potential, with a rather smooth variation of the 
restoring force in all directions. With EDIP, discontinuous variations are 
obtained for displacements along $<110>$ in both glide and shuffle sets. 
In particular, the restoring force along the favored direction, 
$<110>$ in the shuffle set, increases to a large maximum just before 0.5b, 
and then suddenly drops to a symmetric minimum. This sharp behavior is not 
observed in the ab initio curve. The last potential, Tersoff, shows the 
worst results, with discontinuities between 0.4b and 0.6b in the shuffle 
set, so in the range of large deformation, but also for small 
displacements in the glide set.

\section{Conclusion}

We have investigated the properties of bulk silicon submitted to a homogeneous shear, using ab 
initio techniques. It appeared that the shear takes place almost entirely in the shuffle set planes, 
with only slight displacements in the glide set planes. The atomic bonds between atoms on 
both sides of the shuffle set, loose progressively their covalent character until a metallic 
state is established in the shear direction at half the maximum applied shear. Then 
the reverse process is observed, and the perfect diamond crystal structure is recovered. 
We have shown that glide set plays a predominant 
role on the bulk shear stress and that the strains localized in the glide 
set are linear and elastic with respect to the bulk shear stresses, with a 
large shear modulus. At 0~K, silicon can be viewed as formed by a stacking of 
'elastic' glide set planes and 'plastic' shuffle set planes. Our results are confirmed 
by the analysis of GSF energy surfaces and restoring forces, determined with ab initio calculation \cite{Jua96PMA}, 
which suggests that shuffle set is the favored place for the glide event at 0~K. 
The variations of bulk shear stress are similar to the variations of the restoring force in 
the active glide plane. So at 0~K, a correct description of the 
restoring force is a prerequisite to model glide events.

One of the main objective of this work was the determination of the best interatomic potentials in the 
case of largely deformed silicon systems. We have then performed calculations of sheared bulk silicon and 
GSF energy surfaces with SW, Tersoff and EDIP potentials, and compared with ab initio results. 
For sheared bulk silicon we observed that EDIP and SW provided a faithful description of the glide 
event, and strains and stresses analysis showed that the theoretical shear strength is better determined 
with SW. Regarding the GSF energy surfaces and restoring forces, it appears that 
the shuffle set is the preferred place for the glide events at 0 K for all 
the potentials, but the best values of the restoring force ($\tau_{max}$) is  
given by SW. It must also be emphasized that the smoothest description of the glide of one plane 
on another is also obtained with SW. In summary, under large strains, SW seems to be the best potential to 
model qualitatively silicon. 

This work was partially motivated by a previous work on the 
dislocation nucleation process from a surface step under an uniaxial stress \cite{God02SM2}. 
In fact, in that case, a large homogeneous shear strain is present in the atomic structure. 
Our results explain why it is possible to model the nucleation process with the SW potential, whereas 
a potential like Tersoff leads to fracture or local amorphization under stress. 

Finally, one possible explanation for the differences between the potentials, such as 
the discontinuities and local minima on the energy and stress, may come from the cut 
off radius of each potential. In fact, when the atomic structure is largely deformed, the 
number of atoms taken into account 
in the energy calculation may abruptly changes, leading to sharp energy variation.
The smooth behavior of SW may then be explained by its relatively large 
cut off radius of 3.77\AA. Another possible explanation is the simplicity of its functional form 
and the small number of fitted parameters. For Tersoff and EDIP, the functional is more complicated 
with more parameters. While this is required to model properly a large range of experimental 
quantities, a more complex functional may lead to non physical behaviors, when the atomic structure is 
far from the equilibrium state.

\clearpage

\section*{Table captions}

\begin{table}[h] 
\caption{Theoretical shear strengths and strains obtained with 
different potentials.}\label{theo_stress} 
\end{table}

\begin{table}[h] 
\caption{Unstable stacking fault energies $\gamma_{us}$ along relevant Burgers 
vectors (\textbf{b}), for the shuffle and glide planes in (J/m$^{2}$), Unrelaxed 
(U) and Relaxed (R) with atomic and volume 
relaxation. (The $\gamma_{us}$ is not necessarily localized at {\bf f} = b/2.)}\label{GSF} 
\end{table}

\begin{table}[h] 
\caption{Maximum value of the restoring force $\tau_{max}$ for 
the relevant directions (in eV/\AA$^{3}$).}\label{taumax} 
\end{table}

\clearpage

\vspace{4cm}

\begin{table}[h] 
\begin{center} 
\begin{tabular}{cccccc} 
\hline 
\hline 
&\multicolumn{4}{c}{Constant volume}& volume relaxation \\

                 & SW & Tersoff & EDIP & DFT-LDA & DFT-LDA \cite{Ume02MSE} \\ 
\hline 
Theoretical shear strength (GPa)  & 9.6 & 16.7 & 13.9 & 7.95 & 10 \\
Theoretical shear strain (\%)   & 32.7 & 53 & 53 & 24.5 & 30 \\
(\% of the applied strain)      & 27 & 43 & 43 & 20 & 25 \\
\hline 
\hline 
\end{tabular} 
\end{center} 
\end{table}

\vfill\centerline{\bf Table~\ref{theo_stress}}


\vspace{4cm}

\begin{table}[h] 
\begin{center} 
\begin{tabular}{ccccccccc} 
\hline 
\hline 
& \multicolumn{2}{c}{SW} & \multicolumn{2}{c}{Tersoff} & \multicolumn{2}{c}{EDIP} 
& \multicolumn{2}{c}{DFT-LDA \cite{Jua96PMA}} \\
& U & R & U & R & U & R & U & R \\ 
\hline 
$<110>$ shuffle &  1.38 & 0.83 &  2.57 & 1.50 &  2.16 & 1.32 & 1.84  & 1.67 \\ 
$<112>$ glide   &  4.78 & 3.08 &  3.33 & 1.96 &  3.24 & 1.71 & 2.51  & 1.91 \\ 
$<110>$ glide   & 26.09 & 6.21 & 31.19 & 5.27 & 13.43 & 6.14 & 24.71 & $\simeq$5.55 \\ 
\hline \hline 
\end{tabular} 
\end{center} 
\end{table}

\vfill\centerline{\bf Table~\ref{GSF}}


\vspace{4cm}

\begin{table}[h]  
\begin{center} 
\begin{tabular}{ccccc} 
\hline 
\hline 
               & SW & Tersoff & EDIP & DFT-LDA \cite{Jua96PMA}\\ 
\hline 
$<110>$ shuffle & 0.055 & 0.144 & 0.160 & 0.093 \\ 
$<112>$ glide   & 0.299 & 0.535 & 0.322 & 0.174 \\ 
$<110>$ glide   & 0.437 & 0.688 & 1.908 & 0.268 \\ 
\hline 
\hline 
\end{tabular} 
\end{center} 
\end{table}

\vfill\centerline{\bf Table~\ref{taumax}}

\clearpage

\section*{Figure captions}

\begin{figure}[h] 
\caption{Definition of displacements and shear strains shown on a deformed structure (right) compared to 
the perfect lattice (left). In shuffle set planes
the displacement is called D$_{sh}$, and the shear strain 
is defined as $\frac {D_{sh}}{H_{sh}}$, H$_{sh}$ being the 'height' of the shuffle set. 
In glide set planes, similar notations are taken.}\label{strain} 
\end{figure}

\begin{figure}[h] 
\caption{Snapshot of the bulk structure during homogeneous 
shear process. Here, bonds are drawn solely on the criterion of distance and 
are not indicative of a true chemical bonds between atoms.}\label{snapshot} 
\end{figure}

\begin{figure}[h] 
\caption{Upper graph: variation of atomic energy during the shear process (in eV/atom). Lower graphs : 
bulk shear stress $\sigma_{23}$ in GPa (solid line) and shear strain of the 
glide plane multiplied by a shear modulus in GPa (dashed line), for the different 
potentials.}\label{shear} 
\end{figure}

\begin{figure}[h] 
\caption{The left panel shows the displacements in both shuffle (dashed line) and 
glide (dotted line) plane in unit b vs the applied
shear strain. The solid line corresponds to the addition of shuffle and glide displacements.
The right panel shows the shear strain in both shuffle (dashed line) and glide (doted line)
planes vs the applied shear strain. The solid line correspond to the applied strain.}\label{disp} 
\end{figure}

\begin{figure}[h] 
\caption{The fully relaxed GSF energies and the 
corresponding restoring force ($-\tau$) on the 
shuffle (left) and glide (right) set planes (bold lines for $<110>$ and thin 
line for $<112>$ directions) 
for the three interatomic 
potentials.}\label{tau} 
\end{figure}

\clearpage

\ \vspace{2cm}

\begin{center} 
\includegraphics[width=8cm]{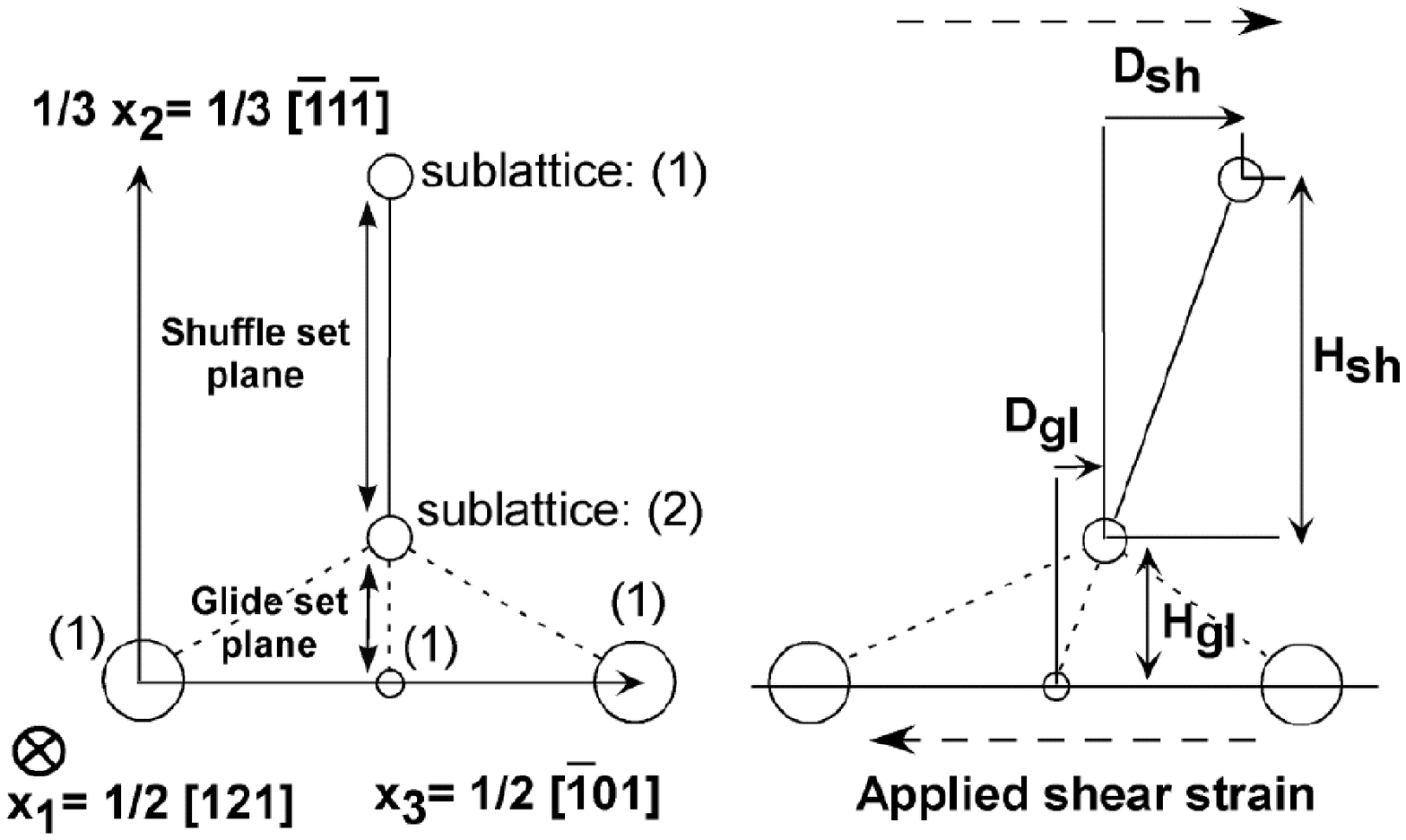} 
\end{center}

\vfill\centerline{\bf Figure~\ref{strain}}

\clearpage

\ \vspace{2cm}

\begin{center} 
\includegraphics[width=8cm]{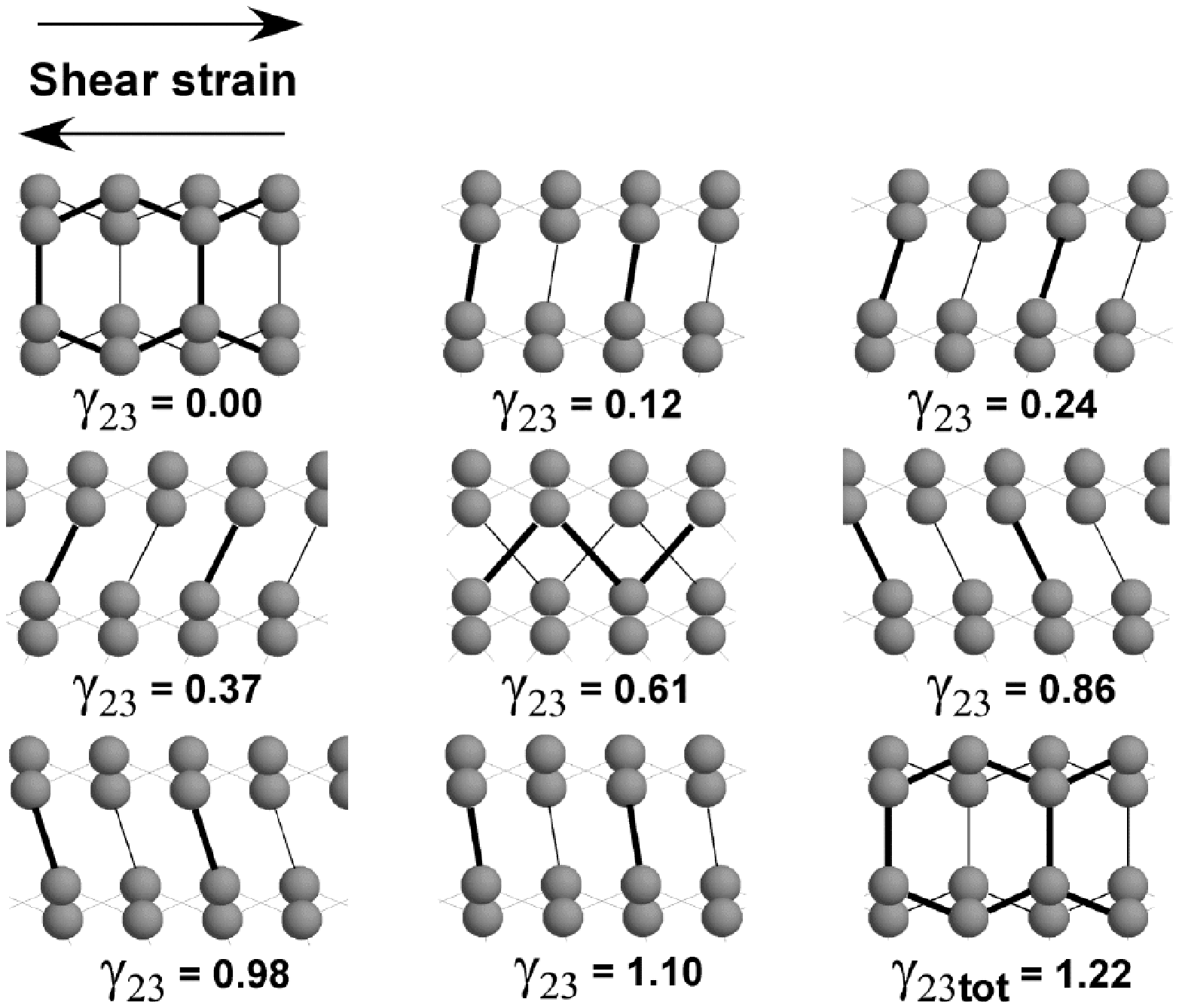} 
\end{center}

\vfill\centerline{\bf Figure~\ref{snapshot}}

\clearpage

\ \vspace{2cm}

\begin{center} 
\includegraphics[width=10cm]{fig3.eps} 
\end{center}

\vfill\centerline{\bf Figure~\ref{shear}}

\clearpage

\ \vspace{2cm}

\begin{center} 
\includegraphics[width=10cm]{fig4.eps} 
\end{center}

\vfill\centerline{\bf Figure~\ref{disp}}

\clearpage

\ \vspace{2cm}

\begin{center} 
\includegraphics[width=14cm]{fig5.eps} 
\end{center}

\vfill\centerline{\bf Figure~\ref{tau}}


\newpage

\section*{References}

\begin{thebibliography}{10}

\bibitem{Obe95PRB}
\"{O}berg S, Sitch P K, Jones R and Heggie M I 1995 {\em Phys. Rev.} B {\bf 51} 13138 

\bibitem{Jus00JPCM}
Justo J F, Fazzio A and Antonelli A 2000 {\em J. Phys.: Condens. Matter} {\bf 12} 10039 

\bibitem{Ewe01JPCM}
Ewels C P, Wilson N T, Heggie M I, Jones R and Briddon P R 2001 
{\em J. Phys.: Condens. Matter} {\bf 13} 8965

\bibitem{Jus01SSC}
Justo J F, Antonelli A and Fazzio A 2001 {\em Solid State Commun.}
{\bf 118} 651

\bibitem{Piz03PM}
Pizzagalli L, Beauchamp P and Rabier J 2003
{\em Phil. Mag.} {\bf 83} 1191

\bibitem{Ras00PMA}
Rasmussen T, Vegge T, Leffers T, Pedersen O B and Jacobsen K W 2000
{\em Phil. Mag. }A {\bf 80} 1273

\bibitem{Bal92PRB}
Balamane H, Halicioglu T and Tiller W A 1992
{\em Phys. Rev. }B {\bf 46} 2250

\bibitem{Jus98PRB}
Justo J F, Bazant M Z, Kaxiras E, Bulatov V V and Yip S 1998
{\em Phys. Rev. }B {\bf 58} 2539

\bibitem{Mor98JJAP}
Moriguchi K and Shintani A 1998
{\em Jpn. J. Appl. Phys.} {\bf 37} 414

\bibitem{God02SM2}
Godet J, Pizzagalli J, Brochard S and Beauchamp P 2002
{\em Scripta Materialia} {\bf 47} 481

\bibitem{Wu01PMA}
Wu R X and Weatherly G C 2001
{\em Phil. Mag. }A {\bf 81} 1489

\bibitem{Arg01SM}
Argon A S and Gally B J 2001
{\em Superlatt. Microstruc.} {\bf 45} 1287

\bibitem{God03UNP}
Godet J, Pizzagalli L, Brochard B and Beauchamp P, to be published.

\bibitem{Due96PML}
Duesbery M S and Jo{\'o}s B 1996
{\em Phil. Mag. Lett.} {\bf 74} 253

\bibitem{Jus01PHY}
Justo J F, Antonelli A and Fazzio A 2001
{\em Physica\ }B {\bf 302-303} 398

\bibitem{Bul01PMA}
Bulatov V V, Justo J F, Cai W, Yip S, Argon A S, Lenosky T,
  de~Koning~M. and Diaz de~la Rubia~T 2001
{\em Phil. Mag. }A {\bf 81} 1257

\bibitem{Rab01MSE}
Rabier J, Cordier P, Demenet J L and Garem H 2001
{\em Mater. Sci. Eng. }A {\bf A309-A310} 74

\bibitem{Per00AM}
P{\'e}rez R and Gumbsch P
{\em Acta Mater.} {\bf 48} 4517

\bibitem{Gal01PMA}
Gally B J and Argon A S 2001
{\em Phil. Mag. }A {\bf 81} 699

\bibitem{Pir01PMA}
Pirouz P, Demenet J L and Hong M H 2001
{\em Phil. Mag. }A {\bf 81} 1207

\bibitem{Hir82WIL}
Hirth J P and Lothe J 1982
{\em Theory of dislocations} (Wiley)

\bibitem{Abinit}
The ABINIT code is a common project of the Universit{\'e} Catholique de
  Louvain, Corning Incorporated, and other contributors (URL
  http://www.abinit.org).

\bibitem{Goe96PRB}
Goedecker S, Teter M and Huetter J 1996
{\em Phys. Rev. },B {\bf 54} 1703

\bibitem{Tro91PRB}
Troulier N and Martins J L 1991
{\em Phys. Rev. }B {\bf 43} 1993

\bibitem{Mon76PRB}
Monkhorst H J and Pack J D 1976
{\em Phys. Rev. }B {\bf 13} 5188

\bibitem{Ume02MSE}
Umeno Y and Kitamura T 2002
{\em Mater. Sci. Eng.} B {\bf B88} 79

\bibitem{Sti85PRB}
Stillinger F H and Weber T A 1985
{\em Phys. Rev. }B {\bf 31} 5262

\bibitem{Ter89PRB}
Tersoff J 1989
{\em Phys. Rev. }B {\bf 39} 5566

\bibitem{Baz97PRB}
Bazant M Z, Kaxiras E and Justo J F 1997
{\em Phys. Rev. }B {\bf 56} 8542

\bibitem{Kax93PRL2}
Kaxiras E and Duesbery M S 1993
{\em Phys. Rev. Lett.} {\bf 70} 3752

\bibitem{Jua96PMA}
Juan Y M and Kaxiras E 1996
{\em Phil. Mag. }A {\bf 74} 1367

\bibitem{Due91EMRS}
Duesbery M S, Michel D J, Kaxiras E and Joos B 1991
{\em Mat. Res. Soc. Symp. Proc.} {\bf 209} 125

\bibitem{Kon98PRB}
de~Koning M, Antonelli A, Bazant M Z and Kaxiras E 1998
{\em Phys. Rev. }B {\bf 58} 12555

\bibitem{Kar97JPCM}
Karki B B, Ackland G J and Crain J 1997
{\em J. Phys.: Condens. Matter} {\bf 9} 8579

\end{thebibliography}
\end{document}